\documentclass[10pt,article]{IEEEtran}
\usepackage{epsf,cite,amsmath,amscd,amssymb,graphicx,latexsym,multicol}
\begin{document}
\bibliographystyle{ieeetr}

\title{{A Non-Cooperative Power Control Game in
Delay-Constrained Multiple-Access Networks}\vspace{2mm}}

\author{\authorblockN{Farhad Meshkati, H. Vincent Poor, and Stuart C. Schwartz\thanks{This
research was supported by the National Science Foundation under
Grant ANI-03-38807.}}\\
\authorblockA{Department of Electrical Engineering\\Princeton University\\
Princeton, NJ 08544 USA\\
Email: \{meshkati, poor, stuart\}@princeton.edu } }

%

\pagestyle{empty}
\maketitle

%
%
%
%
%

\newtheorem{proposition}{Proposition}
\newenvironment{thmproof}[1]
{\noindent\hspace{2em}{\it #1 }}
{\hspace*{\fill}~\QED\par\endtrivlist\unskip}

\centerfigcaptionstrue


\thispagestyle{empty}

\begin{abstract}
A game-theoretic approach for studying power control in
multiple-access networks with transmission delay constraints is
proposed. A non-cooperative power control game is considered in
which each user seeks to choose a transmit power that maximizes
its own utility while satisfying the user's delay requirements.
The utility function measures the number of reliable bits
transmitted per joule of energy and the user's delay constraint is
modeled as an upper bound on the delay outage probability. The
Nash equilibrium for the proposed game is derived, and its
existence and uniqueness are proved. Using a large-system
analysis, explicit expressions for the utilities achieved at
equilibrium are obtained for the matched filter, decorrelating and
minimum mean square error multiuser detectors. The effects of
delay constraints on the users' utilities (in bits/Joule) and
network capacity (i.e., the maximum number of users that can be
supported) are quantified.
\end{abstract}

\section{Introduction}
In wireless networks, power control is used for resource
allocation and interference management. In multiple-access CDMA
systems such as the uplink of cdma2000, the purpose of power
control is for each user terminal to transmit enough power so that
it can achieve the desired quality of service (QoS) without
causing unnecessary interference for other users in the network.
Depending on the particular application, QoS can be expressed in
terms of throughput, delay, battery life, etc. Since in many
practical situations, the users' terminals are battery-powered, an
efficient power management scheme is required to prolong the
battery life of the terminals. Hence, power control plays an even
more important role in such scenarios.

Consider a multiple-access DS-CDMA network where each user wishes
to locally and selfishly choose its transmit power so as to
maximize its utility and at the same time satisfy its delay
requirements. The strategy chosen by each user affects the
performance of other users through multiple-access interference.
There are several questions to ask concerning this interaction.
First of all, what is a reasonable choice of a utility function
that measures energy efficiency and takes into account delay
constraints? Secondly, given such a utility function, what
strategy should a user choose in order to maximize its utility? If
every user in the network selfishly and locally picks its
utility-maximizing strategy, will there be a stable state at which
no user can unilaterally improve its utility (Nash equilibrium)?
If such an equilibrium exists, will it be unique? What will be the
effect of delay constraint on the energy efficiency of the
network?

Game theory is the natural framework for modeling and studying
such a power control problem. Recently, there has been a great
deal of interest in applying game theory to resource allocation is
wireless networks. Examples of game-theoretic approaches to power
control are found in \cite{GoodmanMandayam00,JiHuang98,Saraydar02,
Xiao01,Zhou01,Alpcan,Sung,Meshkati_TCOMM}. In
\cite{GoodmanMandayam00,JiHuang98,Saraydar02, Xiao01,Zhou01},
power control is modeled as a non-cooperative game in which users
choose their transmit powers in order to maximize their utilities.
In \cite{Meshkati_TCOMM}, the authors extend this approach to
consider a game in which users can choose their uplink receivers
as well as their transmit powers. All the power control games
proposed so far assume that the traffic is not delay sensitive.
Their focus is entirely on the trade-offs between throughput and
energy consumption without taking into account any delay
constraints. In this work, we propose a non-cooperative power
control game that does take into account a transmission delay
constraint for each user. Our focus here is on energy efficiency.
Our approach allows us to study networks with both delay tolerant
and delay sensitive traffic/users and quantify the loss in energy
efficiency due to the presence of users with stringent delay
constraints.

The organization of the paper is as follows. In Section
\ref{system model}, we present the system model and define the
users' utility function as well as the model used for
incorporating delay constraints. The proposed power control game
is described in Section \ref{proposed game}, and the existence and
uniqueness of Nash equilibrium for the proposed game is discussed
in Section \ref{Nash equilibrium}. In Section \ref{multiclass}, we
extend the analysis to multi-class networks and derive explicit
expressions for the utilities achieved at Nash equilibrium.
Numerical results and conclusions are given in Sections
\ref{Numerical results} and \ref{conclusions}, respectively.

\section{System Model} \label{system
model}

We consider a synchronous DS-CDMA network with $K$ users and
processing gain $N$ (defined as the ratio of symbol duration to
chip duration). We assume that all $K$ user terminals transmit to
a receiver at a common concentration point, such as a cellular
base station or any other network access point. The signal
received by the uplink receiver (after chip-matched filtering)
sampled at the chip rate over one symbol duration can be expressed
as
\begin{equation}\label{eq1}
   {\mathbf{r}} = \sum_{k=1}^{K} \sqrt{p_k} h_k \ b_k {\mathbf{s}}_k +
   {\mathbf{w}} ,
\end{equation}
where $p_k$, $h_k$, $b_k$ and ${\mathbf{s}}_k$ are the transmit
power, channel gain, transmitted bit and spreading sequence of the
$k^{th}$ user, respectively, and $\mathbf{w}$ is the noise vector
which is assumed to be Gaussian with mean $\mathbf{0}$ and
covariance $\sigma^2 \mathbf{I}$. We assume random spreading
sequences for all users, i.e., $ {\mathbf{s}}_k =
\frac{1}{\sqrt{N}}[v_1 ... v_N]^T$, where the $v_i$'s are
independent and identically distributed (i.i.d.) random variables
taking values in \{$-1,+1$\} with equal probabilities.

\subsection{Utility Function}

To pose the power control problem as a non-cooperative game, we
first need to define a suitable utility function. It is clear that
a higher signal to interference plus noise ratio (SIR) level at
the output of the receiver will result in a lower bit error rate
and hence higher throughput. However, achieving a high SIR level
requires the user terminal to transmit at a high power which in
turn results in low battery life. This tradeoff can be quantified
(as in \cite{GoodmanMandayam00}) by defining the utility function
of a user to be the ratio of its throughput to its transmit power,
i.e.,
\begin{equation}\label{eq2}
   u_k = \frac{T_k}{p_k} \ .
\end{equation}
Throughput is the net number of information bits that are
transmitted without error per unit time (sometimes referred to as
\emph{goodput}). It can be expressed as
\begin{equation}\label{eq3}
   T_k = \frac{L}{M} R_k f(\gamma_k)  ,
\end{equation}
where $L$ and $M$ are the number of information bits and the total
number of bits in a packet, respectively. $R_k$ and $\gamma_k$ are
the transmission rate and the SIR for the $k^{th}$ user,
respectively; and $f(\gamma_k)$ is the ``efficiency function"
which is assumed to be increasing and S-shaped (sigmoidal) with
$f(\infty)=1$. We also require that $f(0)=0$ to ensure that
$u_k=0$ when $p_k=0$. In general, the efficiency function depends
on the modulation, coding and packet size. A more detailed
discussion of the efficiency function can be found in
\cite{Meshkati_TCOMM}. Note that for a sigmoidal efficiency
function, the utility function in \eqref{eq2} is a quasiconcave
function of the user's transmit power. The throughput $T_k$ in
(\ref{eq3}) could also be replaced with any increasing concave
function such as the Shannon capacity formula as long as we make
sure that $u_k = 0$ when $p_k=0$.

Based on (\ref{eq2}) and (\ref{eq3}), the utility function for
user $k$ can be written as
\begin{equation}\label{eq4}
   u_k = \frac{L}{M} R \frac{f(\gamma_k)}{p_k}\ .
\end{equation}
For the sake of simplicity, we have assumed that the transmission
rate is the same for all users, i.e., $R_1 = ... = R_K = R$. All
the results obtained here can be easily generalized to the case of
unequal rates. The utility function in \eqref{eq4}, which has
units of \emph{bits/Joule}, captures very well the tradeoff
between throughput and battery life and is particularly suitable
for applications where energy efficiency is crucial.

\subsection{Delay Constraints}\label{delay constraint}

Let $X$ represent the (random) number of transmissions required
for a packet to be received without any errors. The assumption is
that if a packet has one or more errors, it will be retransmitted.
We also assume that retransmissions are independent from each
other. It is clear that the transmission delay for a packet is
directly proportional to $X$. Therefore, any constraint on the
transmission delay can be equivalently expressed as a constraint
on the number of transmissions. Assuming that the packet success
rate is given by the efficiency function $f(\gamma)$\footnote{This
assumption is valid in many practical systems (see
\cite{Meshkati_TCOMM} for further details).}, the probability that
exactly $m$ transmissions are required for the successful
transmission of the packet is given by
\begin{equation}\label{eq5}
\textrm{Pr}\{X=m\}= f(\gamma) \left( 1-f(\gamma) \right)^{m-1} ,
\end{equation}
and, hence, $\textrm{E}\{X\}=\frac{1}{f(\gamma)}$. We model the
delay requirements of a particular user (or equivalently traffic
type) as a pair $(D,\beta)$, where
\begin{equation}\label{eq6}
\textrm{Pr}\{X\leq D\}\geq \beta .
\end{equation}
In other words, we would like the number of transmissions to be at
most $D$ with a probability larger than or equal to $\beta$. For
example, $(2,0.9)$, i.e., $D=2$ and $\beta=0.9$, implies that 90\%
of the time we need at most two transmissions to successfully
receive a packet. Note that \eqref{eq6} can equivalently be
represented as an upper bound on the delay outage probability,
i.e.,

\begin{equation}
P_{delay\ outage}\triangleq \textrm{Pr}\{X > D\}\leq 1-\beta .
\end{equation}

Based on \eqref{eq5}, the delay constraint in \eqref{eq6} can be
expressed as
$$\sum_{m=1}^D f(\gamma) \left(1-f(\gamma)\right)^{m-1} \geq
\beta ,$$ or
\begin{equation}\label{eq7}
   f(\gamma) \geq \eta(D,\beta) ,
\end{equation}
where
\begin{equation}\label{eq7b}
   \eta(D,\beta)=1-(1-\beta)^{\frac{1}{D}} .
\end{equation}
Here, we have explicitly shown that $\eta$ is a function of $D$
and $\beta$. Since $f(\gamma)$ is an increasing function of
$\gamma$, we can equivalently express \eqref{eq7} as
\begin{equation}\label{eq8}
   \gamma \geq \tilde{\gamma} ,
\end{equation}
where
\begin{equation}\label{eq9}
  \tilde{\gamma}= f^{-1} \left( \eta (D,\beta) \right).
\end{equation}
Therefore, the delay constraint in \eqref{eq6} translates into a
lower bound on the SIR. Since different users could have different
delay requirements, $\tilde{\gamma}$ is user dependent. We make
this explicit by writing
\begin{equation}\label{eq10}
   \tilde{\gamma}_k= f^{-1} \left( \eta_k \right) ,
\end{equation}
where $\eta_k=1-(1-\beta_k)^{\frac{1}{D_k}}$. A more stringent
delay requirement, i.e., a smaller $D$ and/or a larger $\beta$,
will result in a higher value for $\tilde{\gamma}$. Without loss
of generality, we have assumed that all the users in the network
have the same efficiency function. It is straightforward to relax
this assumption.

\section{Power Control Game with Delay Constraints}\label{proposed game}

We propose a power control game in which each user decides how
much power to transmit in order to maximize its own utility and at
the same time satisfy its delay requirements. We have shown in
Section \ref{delay constraint} that the delay requirements of a
user translate into a lower bound on the user's output SIR.
Therefore, each user will seek to maximize its utility while
satisfying its SIR requirement. This can be captured by defining a
\emph{delay-constrained} utility for user $k$ as
\begin{equation}\label{eq11}
   \tilde{u}_k = \left\{%
\begin{array}{ll}
   u_k & \ \ \textrm{if} \ \ \gamma_k \geq \tilde{\gamma}_k \\
   0   & \ \ \textrm{if} \ \ \gamma_k < \tilde{\gamma}_k \\
\end{array}%
\right. ,
\end{equation}
where $u_k$ and $\tilde{\gamma}_k$ are given by \eqref{eq4} and
\eqref{eq10}, respectively.

Let $\tilde{G}=[{\mathcal{K}}, \{A_k \}, \{\tilde{u}_k \}]$ denote
the proposed non-cooperative game where ${\mathcal{K}}=\{1, ... ,
K \}$, and $A_k=[0,P_{max}]$, which is the strategy set for the
$k^{th}$ user. Here, $P_{max}$ is the maximum allowed power for
transmission. We assume that only those users whose delay
requirements can be met are admitted into the network. For
example, for the conventional matched filter, this translates into
having $$\sum_{k=1}^K \frac{1}{1+\frac{N}{\tilde{\gamma}_k}} <1
.$$ This assumption makes sense because admitting a user that
cannot meet its delay requirement only causes unnecessary
interference for other users.

The resulting non-cooperative game can be expressed as the
following maximization problem:
\begin{equation}\label{eq12}
   \max_{p_k} \ \tilde{u}_k  \ \textrm{for} \ \  k=1,...,K ,
\end{equation}
where the $p_k$'s are constrained to be non-negative. The above
maximization can equivalently be written as
\begin{equation}\label{eq13}
   \max_{p_k} \ u_k  \ \ \textrm{subject to} \ \ \gamma_k \geq \tilde{\gamma}_k \ \ \textrm{for} \ \
   k=1,...,K.
\end{equation}
Let us first solve the above maximization by ignoring the
constraints on SIR. For all linear receivers, we have
\begin{equation}\label{eq14}
\frac{\partial \gamma_k}{\partial p_k} = \frac{\gamma_k}{p_k} \ .
\end{equation}
Taking the derivative of $u_k$ with respect to $p_k$ and taking
advantage of \eqref{eq14}, we obtain $$\frac{\partial
u_k}{\partial p_k}= \frac{f'(\gamma_k)}{p_k}\frac{\partial
\gamma_k}{\partial p_k} - \frac{f(\gamma_k)}{p_k^2}=\frac{\gamma_k
f'(\gamma_k) - f(\gamma_k)}{p_k^2} .$$ Therefore, the
unconstrained utility function for user $k$ is maximized when the
user's output SIR is equal to $\gamma^*$, where $\gamma^*$ is the
(positive) solution to
\begin{equation}\label{eq15b}
f(\gamma) = \gamma \ f'(\gamma) .
\end{equation}
It can be shown that for a sigmoidal efficiency function,
$\gamma^*$ always exists and is unique. In addition, for all
$\gamma_k<\gamma^*$, $u_k$ is increasing in $p_k$ and for all
$\gamma_k>\gamma^*$, $u_k$ is decreasing in $p_k$ \cite{Rod03b}.
Therefore, $\tilde{u}_k$ is maximized when user $k$ transmits at a
power level that achieves $\tilde{\gamma}_k^*$ at the output of
the uplink receiver, where
\begin{equation}\label{eq15}
   \tilde{\gamma}_k^*= \max\{\tilde{\gamma}_k,\gamma^*\} .
\end{equation}

In the next section, we investigate the existence and uniqueness
of Nash equilibrium for our proposed game.

\section{Nash Equilibrium for the Proposed Game}\label{Nash
equilibrium}

The Nash equilibrium for the proposed game is a set strategies
(power levels) for which no user can unilaterally improve its own
(delay-constrained) utility function. We now state the following
proposition.

\begin{proposition}\label{prop1}
The Nash equilibrium for the non-cooperative game $\tilde{G}$ is
given by $\tilde{p}_k^*=\min \{p_k^*, P_{max}\}$, for $k=1,
\cdots, K$, where $p_k^*$ is the transmit power that results in an
SIR equal to $\tilde{\gamma}^*$ at the output of the receiver with
$\tilde{\gamma}_k^*= \max\{\tilde{\gamma}_k,\gamma^*\}$.
Furthermore, this equilibrium is unique.
\end{proposition}
\begin{thmproof}{Proof:} Based on the arguments presented in Section
\ref{proposed game}, $\tilde{u}_k$ is maximized when the transmit
power $p_k$ is such that $\gamma_k=\tilde{\gamma}_k^*=
\max\{\tilde{\gamma}_k,\gamma^*\}$. If $\tilde{\gamma}_k$ cannot
be achieved, the user must transmit at maximum power level to
maximize its utility. Let us define $\tilde{p}_{k}$ as the power
level for which the output SIR for user $k$ is equal to
$\tilde{\gamma}_k$. Since user $k$ is admitted into the network
only if it can meet its delay requirements, we have $\tilde{p}_{k}
\leq P_{max}$. In addition, because $\tilde{u}_k=0$ for
$p_k<\tilde{p}_{k}$, there is no incentive for user $k$ to
transmit at a power level smaller than $\tilde{p}_{k}$. Therefore,
we can restrict the set of strategies for user $k$ to
$\left[\tilde{p}_{k} , P_{max}\right]$. In this interval,
$\tilde{u}_k=u_k$ and hence the utility function is continuous and
quasiconcave. This guarantees existence of a Nash equilibrium for
the proposed power control game.

Furthermore, for a sigmoidal efficiency function, $\gamma^*$,
which is the (positive) solution of $f(\gamma) = \gamma \
f'(\gamma)$, is unique and as a result $\tilde{\gamma}_k^*$ is
unique for $k=1,2,...,K$. Because of this and the one-to-one
correspondence between the transmit power and the output SIR, the
Nash equilibrium is unique. \vspace{0.1cm}
\end{thmproof}

The above proposition suggests that at Nash equilibrium, the
output SIR for user $k$ is $\tilde{\gamma}^*_k$, where
$\tilde{\gamma}^*_k$ depends on the efficiency function through
$\gamma^*$ as well as user $k$'s delay constraint through
$\tilde{\gamma}_k$. Note that this result does not depend on the
choice of the receiver and is valid for all linear receivers
including the matched filter, the decorrelator and the (linear)
minimum mean square error (MMSE) detector.

\section{Multi-class Networks}\label{multiclass}

Let us now consider a network with $C$ classes of users. The
assumption is that all the users in the same class have the same
delay requirements characterized by the corresponding $D$ and
$\beta$. Based on Proposition \ref{prop1}, at Nash equilibrium,
all the users in class $c$ will have the same output SIR,
$\tilde{\gamma}^{* (c)}= \max\{\tilde{\gamma}^{(c)},\gamma^*\}$,
where $\tilde{\gamma}^{(c)}=f^{-1} \left( \eta^{(c)} \right)$.
Here, $\eta^{(c)}$ depends on the delay requirements of class $c$,
namely $D^{(c)}$ and $\beta^{(c)}$, through \eqref{eq7b}. The goal
is to quantify the effect of delay constraints on the energy
efficiency of the network or equivalently on the users' utilities.

In order to obtain explicit expressions for the utilities achieved
at equilibrium, we use a large-system analysis similar to the one
presented in \cite{TseHanly99} and \cite{Comaniciu03}. We consider
the asymptotic case in which $K, N \rightarrow \infty $ and
$\frac{K}{N} \rightarrow \alpha < \infty$. This allows us to write
SIR expressions that are independent of the spreading sequences of
the users. Let $K^{(c)}$ be the number of users in class $c$, and
define $\alpha^{(c)}=\lim_{K,N\rightarrow \infty}
\frac{K^{(c)}}{N}$. Therefore, we have $\sum_{c=1}^{C}
\alpha^{(c)} = \alpha$.

It can be shown that for the matched filter (MF), the decorrelator
(DE), and the MMSE detector, the minimum power required by user
$k$ in class $c$ to achieve an output SIR equal to
$\tilde{\gamma}^{* (c)}$ is given by the following equations:
{\small {\begin{eqnarray} 
    &p_k^{MF}& = \frac{1}{h_k^2} \frac{ \tilde{\gamma}^{* (c)} \sigma^2 }{ 1 - \sum_{c=1}^{C}
    \alpha^{(c)} \tilde{\gamma}^{* (c)} } \nonumber\\ && \hspace{2.5cm} \ {\textrm{for}} \ \ \sum_{c=1}^{C}
    \alpha^{(c)} \tilde{\gamma}^{* (c)} < 1, \ \label{eq20-1}\\
    &&\nonumber\\
   &p_k^{DE}& = \frac{1}{h_k^2} \frac{\tilde{\gamma}^{* (c)} \sigma^2}{1 - \alpha}
 \ \ \ \ \ \ \ {\textrm{for}} \ \ \alpha < 1, \label{eq20-2}\\
\textrm{and} \nonumber\\
    &p_k^{MMSE}& = \frac{1}{h_k^2} \frac{\tilde{\gamma}^{* (c)} \sigma^2}{1-\sum_{c=1}^{C} \alpha^{(c)} \frac{\tilde{\gamma}^{*
(c)}}{1+\tilde{\gamma}^{* (c)}} } \nonumber\\ && \hspace{2cm}
{\textrm{for}} \ \ \sum_{c=1}^{C} \alpha^{(c)}
\frac{\tilde{\gamma}^{* (c)}}{1+\tilde{\gamma}^{* (c)}}< 1.
\label{eq20-3}
\end{eqnarray} }}
Note that we have implicitly assumed that $P_{max}$ is
sufficiently large so that the target SIRs (i.e.,
$\tilde{\gamma}^{* (c)}$'s) can be achieved by all users.
Furthermore, since $\tilde{\gamma}^{* (c)} \geq
\tilde{\gamma}^{(c)}$ for $c=1,\cdots, C$, we have $\tilde{u}_k =
u_k = \frac{L}{M} R \frac{f(\tilde{\gamma}^{* (c)})}{p_k}$.
Therefore, for the matched filter, the decorrelator, and the MMSE
detector, the utilities achieved at the Nash equilibrium are given
by
{\small{\begin{eqnarray} 
    &\tilde{u}_k^{MF}& = \frac{L R}{M \sigma^2} h_k^2\left( 1 - \sum_{c=1}^{C}
    \alpha^{(c)} \tilde{\gamma}^{* (c)}\right) \frac{ f(\tilde{\gamma}^{* (c)})}{ \tilde{\gamma}^{* (c)}} \nonumber\\
    && \hspace{2.5cm} {\textrm{for}} \ \ \sum_{c=1}^{C} \alpha^{(c)} \tilde{\gamma}^{* (c)} < 1, \
    \label{eq21-1}\\ &&\nonumber\\
   &\tilde{u}_k^{DE}& = \frac{L R}{M \sigma^2} h_k^2 \left( 1-\sum_{c=1}^{C}
    \alpha^{(c)} \right)  \frac{ f(\tilde{\gamma}^{* (c)})}{ \tilde{\gamma}^{* (c)}} \nonumber\\
    && \hspace{3cm} {\textrm{for}} \ \ \  \sum_{c=1}^{C} \alpha^{(c)}  < 1, \
    \label{eq21-2}\\
\textrm{and} \nonumber\\
  &\tilde{u}_k^{MMSE}& = \frac{L R}{M \sigma^2} h_k^2 \left(1-\sum_{c=1}^{C} \alpha^{(c)} \frac{\tilde{\gamma}^{* (c)}}{1+\tilde{\gamma}^{* (c)}} \right)
  \frac{f(\tilde{\gamma}^{* (c)})}{ \tilde{\gamma}^{* (c)}} \nonumber\\ &&\hspace{2cm} {\textrm{for}} \ \ \sum_{c=1}^{C} \alpha^{(c)}
  \frac{\tilde{\gamma}^{* (c)}}{1+\tilde{\gamma}^{* (c)}}< 1 . \label{eq21-3}
\end{eqnarray} }}
Note that, based on the above equations, we have
${\tilde{u}_k^{MMSE}>\tilde{u}_k^{DE}>\tilde{u}_k^{MF}}$. This
means that the MMSE reciever achieves the highest utility as
compared to the decorrelator and the matched filter. Also, the
network capacity (i.e., the number of users that can be admitted
into the network) is the highest when the MMSE detector is used.
For the specific case of no delay constraints, $\tilde{\gamma}^{*
(c)}=\gamma^*$ for all $c$ and \eqref{eq21-1}--\eqref{eq21-3}
reduce to
{\small{\begin{eqnarray} 
&u_k^{MF}&=\frac{L R}{M \sigma^2} h_k^2\left( 1 - \alpha
\gamma^*\right) \frac{ f({\gamma}^{*})}{ {\gamma}^{*}}
    \ \ {\textrm{for}} \ \alpha \gamma^* < 1, \ \label{eq22-1} \\&&\nonumber\\
&u_k^{DE}&=\frac{L R}{M \sigma^2} h_k^2 \left( 1-\alpha  \right)
\frac{ f({\gamma}^{*})}{{\gamma}^{*}}
   \ \ {\textrm{for}} \ \ \alpha  < 1, \ \label{eq22-2} \\
\textrm{and} \nonumber \\
&u_k^{MMSE}&=\frac{L R}{M \sigma^2} h_k^2 \left(1- \alpha
\frac{{\gamma}^{*}}{1+{\gamma}^{*}} \right)
  \frac{f({\gamma}^{* })}{ {\gamma}^{*}} \nonumber \\
   && \hspace{3.2cm} {\textrm{for}} \ \ \alpha \frac{{\gamma}^{*}}{1+{\gamma}^{*}}< 1 . \label{eq22-3}
\end{eqnarray} }}
Comparing \eqref{eq21-1}--\eqref{eq21-3} with
\eqref{eq22-1}--\eqref{eq22-3}, we observe that the presence of
users with stringent delay requirements results not only in a
reduction in the utilities of those users but also a reduction in
the utilities of other users in the network. A stringent delay
requirement results in an increase in the user's target SIR
(remember $ \tilde{\gamma}_k^*=
\max\{\tilde{\gamma}_k,\gamma^*\}$). Since
$\frac{f(\gamma)}{\gamma}$ is maximum when $\gamma=\gamma^*$, a
target SIR larger than $\gamma^*$ results in a reduction in the
utility of the corresponding user. In addition, because of the
higher target SIR for this user, other users in the network
experience a higher level of interference and hence are forced to
transmit at a higher power which in turn results in a reduction in
their utilities (except for the decorrelator, in which case the
multiple-access interference is completely removed). Also, since $
\tilde{\gamma}_k^*\geq \gamma^*$ and $\sum_{c=1}^{C}
\alpha^{(c)}=\alpha$, the presence of delay-constrained users
causes a reduction in the system capacity (again, except for the
decorrelator). Through \eqref{eq21-1}--\eqref{eq21-3}, we have
quantified the loss in the utility (in bits/Joule) and in network
capacity due to users' delay constraints for the matched filter,
the decorrelator and the MMSE receiver. The sensitivity of the
loss to the delay parameters (i.e., $D$ and $\beta$) depends on
the efficiency function, $f(\gamma)$.

\section{Numerical Results}\label{Numerical results}

Let us consider the uplink of a DS-CDMA system with processing
gain 100. We assume that each packet contains 100 bits of
information and no overhead (i.e., $L=M=100$). The transmission
rate, $R$, is $100Kbps$ and the thermal noise power, $\sigma^2$,
is $5\times 10^{-16}Watts$.  A useful example for the efficiency
function is $f(\gamma)= (1- e^{-\gamma})^M$. This serves as an
approximation to the packet success rate that is very reasonable
for moderate to large values of $M$ \cite{Saraydar02}. We use this
efficiency function for our simulations. Using this, with $M=100$,
the solution to (\ref{eq15b}) is $\gamma^*=6.48 = 8.1dB$.

\begin{figure}
\begin{center}
\leavevmode \hbox{\epsfysize=5cm \epsfxsize=7cm
\epsffile{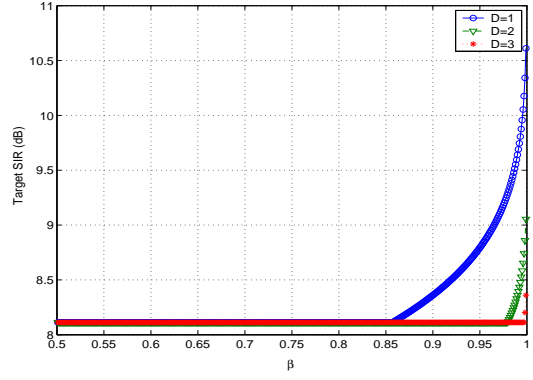}}\vspace{-0.2cm}
\end{center}
\caption{Target SIR, $\tilde{\gamma}^*$, as a Function of $\beta$
for $D=$1,2, and 3.} \label{fig1}
\end{figure}

Fig. \ref{fig1} shows the target SIR as a function of $\beta$ for
$D=1,2,$ and 3. It is observed that, as expected, a more stringent
delay requirement (i.e., a higher $\beta$ and/or a lower $D$)
results in a higher target SIR.

We now consider a network where the users can be divided into two
classes: delay sensitive (class $A$) and delay tolerant (class
$B$). For users in class $A$, we choose $D_A=1$ and $\beta_A=0.99$
(i.e., delay sensitive). For users in class $B$, we let $D_B=3$
and $\beta_B=0.90$ (i.e., delay tolerant). Based on these choices,
$\tilde{\gamma}^*_A=9.6 dB$ and $\tilde{\gamma}^*_B=\gamma^*=8.1
dB$. Without loss of generality and to keep the comparison fair,
we also assume that all the users are 100 meters away from the
uplink receiver. The system load is assumed to be $\alpha$ (i.e.,
$\frac{K}{N}=\alpha$) and we let $\alpha_A$ and $\alpha_B$
represent the load corresponding to class $A$ and class $B$,
respectively, with ${\alpha_A + \alpha_B =\alpha}$.

\begin{figure}
\begin{center}
\leavevmode \hbox{\epsfysize=5cm \epsfxsize=7cm
\epsffile{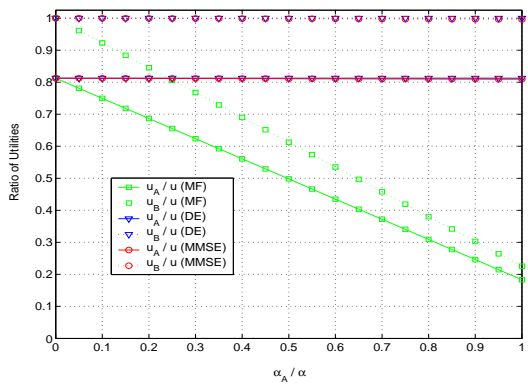}}\vspace{-0.2cm}
\end{center}
\caption{Loss in Utility Due to Presence of Users with Stringent
Delay Requirements ($\alpha=0.1$)} \label{fig2}
\end{figure}

We first consider a lightly loaded network with $\alpha=0.1$ (see
Fig. \ref{fig2}). To demonstrate the performance loss due to the
presence of users with stringent delay requirements (i.e., class
$A$), we plot $\frac{u_A}{u}$ and $\frac{u_B}{u}$ as a function of
the fraction of the load corresponding to class $A$ users (i.e.,
$\frac{\alpha_A}{\alpha}$). Here, $u_A$ and $u_B$ are the
utilities of users in class $A$ and class $B$, respectively, and
$u$ represents the utility of the users if they all had loose
delay requirements which means $\tilde{\gamma}^*_k=\gamma^*$ for
all $k$. Fig. \ref{fig2} shows the loss for the matched filter,
the decorrelator, and the MMSE detector. We observe from the
figure that for the matched filter both classes of users suffer
significantly due to the presence of delay sensitive traffic. For
example, when half of the users are delay sensitive, the utilities
achieved by class $A$ and class $B$ users are, respectively, 50\%
and 60\% of the utilities for the case of no delay constraints.
For the decorrelator, only class $A$ users suffer and the
reduction in utility is smaller than that of the matched filter.
For the MMSE detector, the reduction in utility for class $A$
users is similar to that of the decorrelator, and the reduction in
utility for class $B$ is negligible.

\begin{figure}
\begin{center}
\leavevmode \hbox{\epsfysize=5cm \epsfxsize=7cm
\epsffile{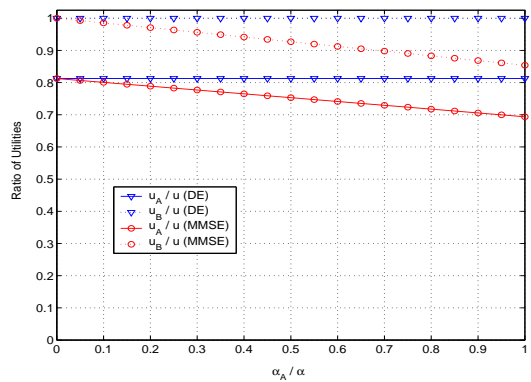}}\vspace{-0.2cm}
\end{center}
\caption{Loss in Utility Due to Presence of Users with Stringent
Delay Requirements ($\alpha=0.9$)} \label{fig3}
\end{figure}

We repeat the experiment for a highly loaded network with
$\alpha=0.9$ (see Fig. \ref{fig3}). Since the matched filter
cannot handle such a significant load, we have shown the plots for
the decorrelator and MMSE detector only. We observe from
Fig.~\ref{fig3} that because of the higher system load, the
reduction in the utilities is more significant for the MMSE
detector compared to the case of $\alpha=0.1$. It should be noted
that for the decorrelator the reduction in utility of class $A$
users is independent of the system load. This is because the
decorrelator completely removes the multiple-access interference.

It should be further noted that in Figs.~\ref{fig2}~and~\ref{fig3}
we have only plotted the ratio of the utilities (not the actual
values). As discussed in Section \ref{multiclass}, the achieved
utilities for the MMSE detector are larger than those of the
decorrelator and the matched filter.

\section{Conclusions}\label{conclusions}

We have proposed a game-theoretic approach for studying power
control in multiple-access networks with (transmission) delay
constraints. We have considered a non-cooperative game where each
user seeks to choose a transmit power that maximizes its own
utility while satisfying the user's delay requirements. The
utility function measures the number of reliable bits transmitted
per joule of energy. We have modeled the delay constraint as an
upper bound on the delay outage probability. We have derived the
Nash equilibrium for the proposed game and have shown that it is
unique. The results are applicable to all linear receivers. In
addition, we have used a large-system analysis to derive explicit
expressions for the utilities achieved at equilibrium for the
matched filter, decorrelator and MMSE detector. The reductions in
the users' utilities (in bits/Joule) and network capacity due to
the presence of users with stringent delay constraints have been
quantified.
%
%
%


\begin{thebibliography}{10}

\bibitem{GoodmanMandayam00}
D.~J. Goodman and N.~B. Mandayam, ``Power control for wireless
data,'' {\em
  IEEE Personal Communications}, vol.~7, pp.~48--54, April 2000.

\bibitem{JiHuang98}
H.~Ji and C.-Y. Huang, ``Non-cooperative uplink power control in
cellular radio
  systems,'' {\em Wireless Networks}, vol.~4, pp.~233--240, April 1998.

\bibitem{Saraydar02}
C.~U. Saraydar, N.~B. Mandayam, and D.~J. Goodman, ``Efficient
power control
  via pricing in wireless data networks,'' {\em IEEE Transactions on
  Communications}, vol.~50, pp.~291--303, February 2002.

\bibitem{Xiao01}
M.~Xiao, N.~B. Shroff, and E.~K.~P. Chong, ``Utility-based power
control in
  cellular wireless systems,'' {\em Proceedings of the Annual Joint Conference
  of the IEEE Computer and Communications Societies (INFOCOM)}, pp.~412--421,
  AK, USA, April 2001.

\bibitem{Zhou01}
C.~Zhou, M.~L. Honig, and S.~Jordan, ``Two-cell power allocation
for wireless
  data based on pricing,'' {\em Proceedings of the $39^{th}$ Annual Allerton
  Conference on Communication, Control, and Computing}, Monticello, IL, USA,
  October 2001.

\bibitem{Alpcan}
T.~Alpcan, T.~Basar, R.~Srikant, and E.~Altman, ``{CDMA} uplink
power control
  as a noncooperative game,'' {\em Proceedings of the $40^{th}$ {IEEE}
  Conference on Decision and Control}, pp.~197--202, Orlando, FL, USA, December
  2001.

\bibitem{Sung}
C.~W. Sung and W.~S. Wong, ``A noncooperative power control game
for multirate
  {CDMA} data networks,'' {\em IEEE Transactions on Wireless Communications},
  vol.~2, pp.~186--194, January 2003.

\bibitem{Meshkati_TCOMM}
F.~Meshkati, H.~V. Poor, S.~C. Schwartz, and N.~B. Mandayam, ``A
utility-based
  appraoch to power control and receiver design in wireless data networks.'' To
  appear in {\it{IEEE Transactions on Communications}}.

\bibitem{Rod03b}
V.~Rodriguez, ``An analytical foundation for resource management
in wireless
  communication,'' {\em Proceedings of the {IEEE} Global Telecommunications
  Conference}, pp.~898--902, San Francisco, CA, USA, December 2003.

\bibitem{TseHanly99}
D.~N.~C. Tse and S.~V. Hanly, ``Linear multiuser receivers:
Effective
  interference, effective bandwidth and user capacity,'' {\em IEEE Transactions
  on Information Theory}, vol.~45, pp.~641--657, March 1999.

\bibitem{Comaniciu03}
C.~Comaniciu and H.~V. Poor, ``Jointly optimal power and admission
control for
  delay sensitive traffic in {CDMA} networks with {LMMSE} receivers,'' {\em
  IEEE Transactions on Signal Processing, Special Issue on Signal Processing in
  Networking}, vol.~51, pp.~2031--2042, August 2003.

\end{thebibliography}
\end{document}